\documentclass[12pt]{iopart}
\usepackage{iopams}
\usepackage{color}
\usepackage{bbold}  
\usepackage{graphicx}
\begin{document}

\title[Experimental quantum cosmology in time-dependent optical media.]{Experimental quantum cosmology in time-dependent optical media.}

\author{N. Westerberg$^1$, S. Cacciatori$^2$, F. Belgiorno$^3$, F. Dalla Piazza$^4$ and D. Faccio$^1$}

\address{$^1$ Institute of Photonics and Quantum Sciences, School of Engineering and Physical Sciences, Heriot-Watt University, EH14 4AS Edinburgh, UK.\\
$^2$ Department of Physics and Mathematics, Universit\`a dell’Insubria, Via Valleggio 11, 22100 Como, Italy.
$^3$ Dipartimento di Matematica, Politecnico di Milano, Piazza Leonardo 32, 20133 Milano, Italy.
$^4$ Dipartimento di Matematica, Universit\`a ``La Sapienza'',  Piazzale A. Moro 2, I-00185 Rome, Italy}
\ead{d.faccio@hw.ac.uk}
\begin{abstract}
It is possible to construct artificial spacetime geometries for light by using intense laser pulses that modify the spatiotemporal properties of an optical medium. Here we theoretically investigate experimental possibilities for studying spacetime metrics of the form $\textrm{d}s^2=c^2\textrm{d}t^2-\eta(t)^2\textrm{d}x^2$. By tailoring the laser pulse shape and medium properties, it is possible to create a refractive index variation $n=n(t)$ that can be identified with $\eta(t)$. Starting from a perturbative solution to a generalised Hopfield model for the medium described by an $n=n(t)$ we provide estimates for the number of photons generated by the time-dependent spacetime. The simplest example is that of a uniformly varying $\eta(t)$ that therefore describes the Robertson-Walker metric, i.e. a cosmological expansion. The number of photon pairs generated in experimentally feasible conditions appears to be extremely small. However, large photon production can be obtained by periodically modulating the medium and thus resorting to a resonant enhancement similar to that observed in the dynamical Casimir effect. Curiously, the spacetime metric in this case closely resembles that of a gravitational wave. Motivated by this analogy we show that a periodic gravitational wave can indeed act as an amplifier for photons. The emission for an actual gravitational wave will be very weak but should be readily observable in the laboratory analogue.

\end{abstract}

\submitto{\NJP}
\maketitle

\section{Introduction.}
\begin{figure}
\centering \includegraphics[width=8cm]{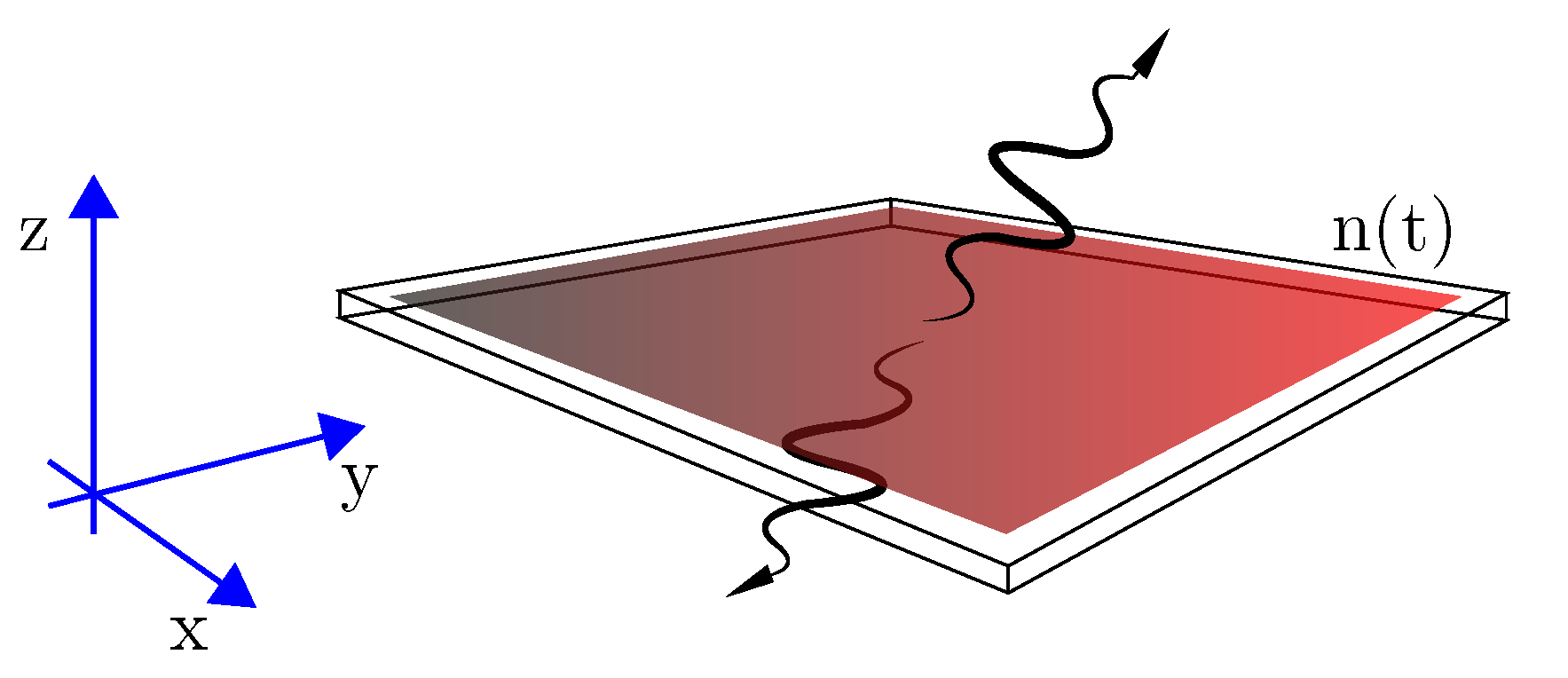}
\caption{\label{fig:layout}  Schematical layout of the system analysed in this work. A thin film of a transparent nonlinear material, e.g. graphene, with thickness $L$ along the z-axis is subject to a laser pulse or series of laser pulses directed along the z-axis and uniformly illuminate the thin film.  The nonlinear response excited by the laser beam creates a time-dependent refractive index $n(t)$. The interaction of the time varying $n(t)$ with the surrounding background vacuum states will lead to the emission of photon pairs.  }
\end{figure}

Analogue gravity is the study of gravitational systems, or more in general of curved spacetimes by means of analogue systems whose kinematics, although based on different underlying physical equations (i.e. dynamics), are governed by identical or similar spacetime metrics \cite{living,book1,book2}. So, whilst it is not possible to use these systems to mimic for example the precise dynamics and interaction between gravitating bodies, it is possible to reproduce the physics of phenomena that rely solely on the specific shape of the spacetime metric. The most studied phenomenon in this sense is without doubt Hawking radiation, i.e. the emission of photons excited out of the vacuum state in the proximity of an event horizon. Originally predicted by S. Hawking to occur for static black holes \cite{haw-nature}, this prediction was later extended by W. Unruh to analogue systems \cite{unruh}, i.e. to acoustic waves propagating in a fluid with a flow that reproduces the same flow of space falling into a black hole. Since this seminal work, many different proposals have been theoretically developed relying on system as varied as Bose-Einstein Condensates, gravity waves in water, microwave radiation in waveguides and ultrashort laser pulse propagation in optical media thus highlighting the ubiquity and general nature of Unruh's idea (see e.g. \cite{living,book1,book2} for an overview). However, it is only in recent years that some of these ideas reached experimental maturity with the first serious attempts to actually create artificial horizons and observe Hawking radiation \cite{belgiorno-prl,stein1,faccioCP}, including the actual realisation of Unruh's original idea of a horizon in a flowing fluid \cite{unruh2}. \\
Clearly, the same rationale of reproducing a spacetime metric and therefore the kinematics with any related quantum behaviour applies not only to black holes but also to other scenarios. Examples are the Unruh effect, a cosmological expansion \cite{visser-cosmo}, metric signature changes \cite{smol1} or, going beyond cosmology, the dynamical Casimir effect \cite{delsing}.\\ Whilst there are many possible different systems in which to build these analogues, here we focus attention on the specific case of the spacetime metric induced by an ultrashort and intense laser pulse propagating through a thin film of material. Leonhardt and co-workers first proposed a similar system, a laser pulse propagating through an optical fibre, as a way to recreate a horizon \cite{philbin-science}. At sufficiently high intensities the laser pulse will excite a nonlinear response in the medium (described by the nonlinear index $n_2$)  that in turn creates a local variation of the refractive index, $n_0$, that can be written as $n(z-vt)=n_0+n_2I(z-vt)$, where $z$ is the propagation direction and $v$ the propagation velocity of a pulse with intensity $I(z-vt)$. The horizon is generated in the comoving frame of the laser pulse where the local increase in refractive index $\delta n=n_2 I$ can slow down a co-propagating light pulse and effectively block it \cite{faccioCP}. Theoretical analysis of this system shows that it is described by the Gordon metric $\textrm{d}s^2=(c/n)^2\textrm{d}t^2-\textrm{d}z^2$ \cite{cacciatoriNJP}, which may be re-written in a similar form to the Painlev\`e-Gullstrant metric for a black hole, $\textrm{d}s^2=c^2\textrm{d}t^2-(\textrm{d}z-v\textrm{d}t)^2$ \cite{belgiornoPRD}. Some simplifying assumptions are made in this derivation, e.g. we neglect optical dispersion and also work in the geometrical optics limit so that the magnetic permeability $\mu$ is constant and the optical analogue is created by acting only upon the dielectric permeability $\varepsilon=n^2$. Nevertheless, numerical simulations based for example of the direct solution of the full Maxwell equations verify the behaviour of the laser induced $\delta n$ in terms of creating a blocking horizon and also the scattering of input waves into two output modes that can be identified with the outgoing positive and infalling negative-energy Hawking modes generated at the horizon \cite{petev}.   \\
Here we expand on this system based on optical nonlinearity to examine a slightly different class of spacetime metrics, namely metrics in which $n=n(t)$, i.e. a medium in which the refractive index varies as a whole. Maxwell's equations are conformally invariant so that we may transform the corresponding spacetime metric into the form
\begin{equation}\label{eq_n}
\textrm{d}s^2=c^2\textrm{d}t^2-n(t)^2\textrm{d}x^2
\end{equation}
where $x$ is a transverse spatial coordinate.  This metric encompasses a large class of different cosmological settings depending on the specific form we choose for $n(t)$, ranging from the Robertson-Walker metric \cite{Ginis} for a cosmological expansion to gravitational waves.\\
The physical system we are studying is schematically depicted in Fig.~\ref{fig:layout}: a thin, square film of a transparent nonlinear material, e.g. graphene is subject to a laser pulse or series of laser pulses directed along the z-axis that uniformly illuminate the thin film.  The nonlinear response excited by the laser beam creates the time-dependent refractive index $n(t)$. \\
In the following we will first describe the model that we use to study photon production from the vacuum state in the presence of the time-varying refractive index. We then apply this model to two specific cases represented by the metric (\ref{eq_n}), namely a cosmological expansion (or contraction) and a periodic spacetime expansion-contraction. The latter metric may be likened to a gravitational wave and, motivated by this analogy, we  show that indeed a gravitational wave may be expected to act as an amplifier for photons.\\

\section{Numerical model}
The main objective of this work is to provide a prediction for the number of photons that will be excited from the vacuum state due to a generic $n=n(t)$ time variation of the refractive index of a given medium. 
In order to do this, we rely on a very general framework based on the Hopfield model that allows us to describe the medium as a collection of oscillators and from this, after performing a pertubrative expansion under the assumption that the $\delta n$ amplitude is small with respect to the background, constant index $n_0$ and therefore evaluate the number of photons in the system after the refractive index change with respect to the initial, stationary input state. The full details of this model are given in Ref.~\cite{hopfield} to which we refer the reader and here we just summarise the main equations.\\
The medium response is described via a Hopfield model as a set of harmonic oscillators that are coupled to the quantised electromagnetic field by a coupling constant.  The full response of the actual medium is described by the coupled dynamics of the bare electromagnetic field with the medium fields. The stationary (non time-varying medium) is described by a Hamiltonian that contains the medium polarization as sum over $N$ resonant oscillator contributions that correspond to the physical resonances that characterise all dispersive media and fully determine the dispersive properties. The medium is therefore characterised by a generic dispersion relation of the form
 \begin{equation}\label{eq_chi}
c^2k^2=\omega^2\left[1 + \sum_{l=0}^N \frac{\chi_l}{\omega^2_l-\omega^2} \right].
\end{equation}
Most dielectric media typically have one or two poles located in the ultraviolet or deep-ultraviolet part of the spectrum (e.g. located at wavelengths shorter than 100-200 nm) and an additional resonance in the mid or far-infrared region (e.g. located at wavelengths longer than 5-10 $\mu$m).  In the following we will refer to a simplified model material in which we consider only a single resonance [$N=0$ in Eq.~(\ref{eq_chi})]. This is justified in light of the fact that we will be examining the behaviour of a limited spectral range in the visible and near infrared region that is indeed dominated in most media of interest by the UV resonance. Moreover, although it is possible to extend the Hopfield model to account for both dispersive and absorptive properties of the medium, here we only consider the dispersive component and neglect any effects due to absorption, a justifiable approximation as long as the electromagnetic field frequencies involved in the interactions are far from the resonance frequencies $\omega_l$. This is readily verified in common experimental conditions - for example, fused silica glass exhibits a resonance in the UV region located at $\sim200$ nm but is essentially transparent to wavelengths longer than 300 nm.\\
Spacetime distortions in the medium are then modelled as a perturbation to the $\chi_0$ parameter, i.e. we use $\chi(x,t)=\chi_0+\delta\chi(x,t)$. The spacetime perturbation is related to the laser pulse induced $\delta n$ by the relation $\delta\chi(x,t)=2(\omega_0^2-\omega^2)n_0\delta n(x,t)$. Therefore, by controlling the laser pulse and the interaction geometry with the medium, we have direct control over the  $\delta\chi(x,t)$ and hence an experimental handle with which to control the photon emission from the modulated medium. \\
In the perturbative limit the number of photon pairs excited by the spacetime varying medium is determined by the scattering matrix $S\simeq\mathbb{1}-(i/\hbar)\int \textrm{d}x\textrm{d}t\delta H(x,t)$, where $\delta H(x,t)$ is the perturbation to the background Hamiltonian. From this it is possible to calculate the number of  photon pairs emitted per unit solid angle, with frequencies $\omega_1$ and $\omega_2$, from the relation
 \begin{equation}\label{eq_hop}
\frac{dN}{dt} = f(\omega_1,\omega_2)\times\left| \widehat{\delta\chi}(\omega_1+\omega_2,k_1+k_2) \right|^2 \textrm{d}\omega_1\textrm{d}\omega_2
\end{equation}
where $\widehat{\delta\chi}$ denotes the Fourier transform of ${\delta\chi}$ and $f(\omega_1,\omega_2)$ is a function of the photon pair frequencies
 \begin{equation}\label{eq_hop2}
f(\omega_1,\omega_2)=\frac{1}{4(2\pi c)^6}\frac{\omega_1^2\omega_2^2(\omega_1\omega_2+\omega_0)^2}{(\omega_0^2-\omega_1^2)^2(\omega_0^2-\omega_2^2)^2}\omega_1n(\omega_1)\omega_2n(\omega_2).
\end{equation}
The main result of this model is therefore that the number of photon pairs depends essentially on the Fourier transform of the refractive index perturbation, parametrized in the model by the resonance oscillator strength $\delta\chi$: it is $\widehat{\delta\chi}$ that determines the precise emission pattern and number of the photon pairs \cite{hopfield}.\\
In the following we proceed to show some specific examples in which this model predicts the generation of photon pairs in time varying spacetimes of the form given in Eq.~(\ref{eq_n}).  

\section{Photon production from an artificial expanding (contracting) universe.} 
Previous studies have outlined how the metric Eq.~(\ref{eq_n}) may be identified with the one dimensional version of the Robertson-Walker metric that governs the kinematics in an expanding or contracting Universe. These studies examined the behaviour of media that reproduce this metric from a classical perspective, e.g. an analysis was given of the red-shift imparted upon a classical probe laser beam. Here we extend this analysis to incorporate quantum effects and estimate the rate of photon pair production from an artificial expanding universe.\\
\begin{figure}
\centering \includegraphics[width=10cm]{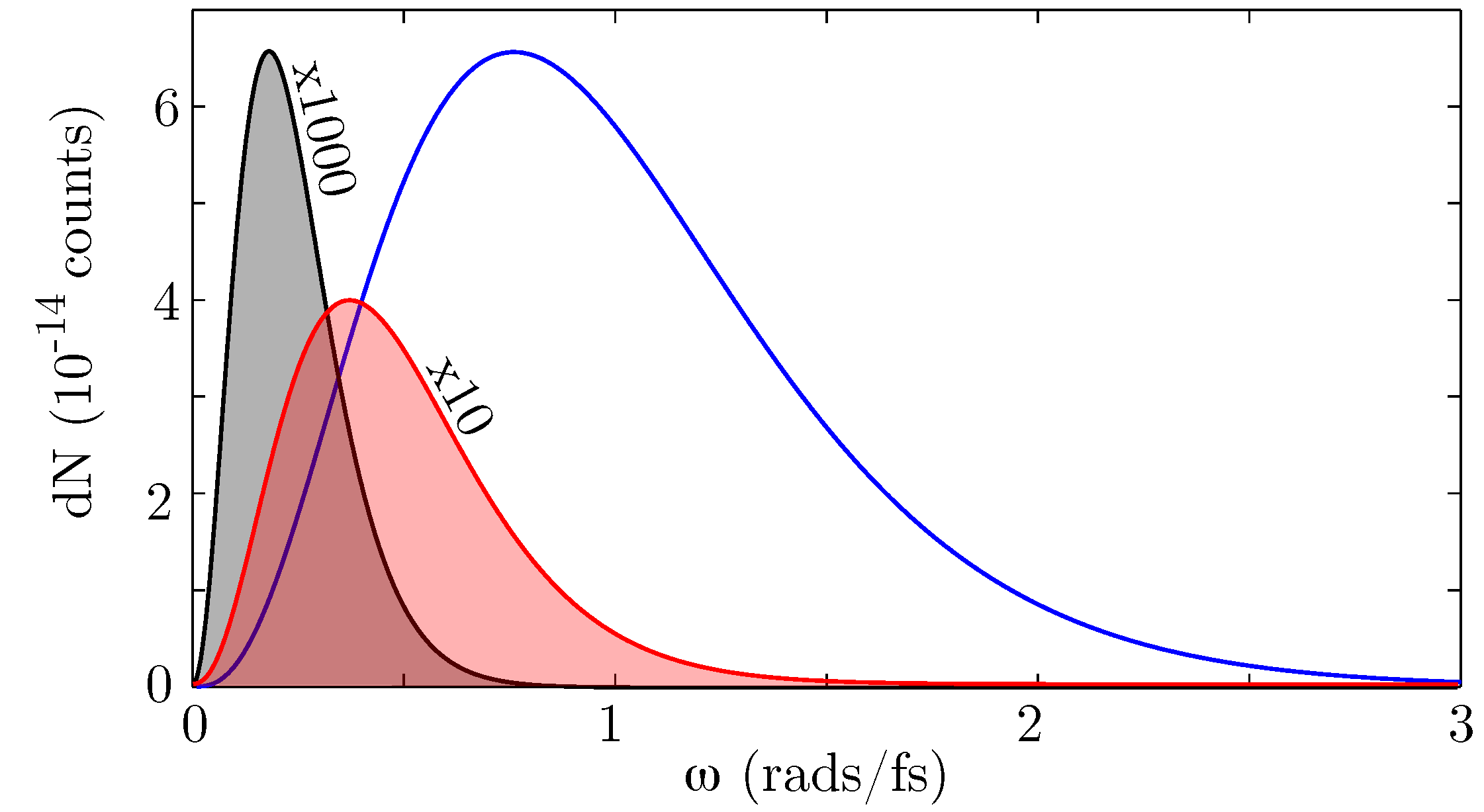}
\caption{\label{fig:expansion}  Numerical estimates of the emitted photon numbers for $n(t)=\textrm{tanh}(\alpha t)$, with $\alpha=20 \times 10^{13}$ 1/s (blue curve), $\alpha=40 \times 10^{13}$ (red shaded curve, multiplied by 10), $\alpha=80\times10^{13}$ (black shaded curve, multiplied by 1000). }
\end{figure}
 From an experimental perspective, the challenge is to create a medium in which the expansion is indeed constant, i.e. such that $n(t)$ is a continuously increasing or decreasing function of time. Any attempt to obtain this over long time scales will lead to either huge or vanishing refractive indices, respectively and are therefore likely to fail. Moreover, as we will see below, the rate at which the artificial universe expands or contracts, i.e. the time gradient of $n(t)$ is also important. However, the expansion or contraction of the medium does not need to proceed indefinitely and the only requirement is that it is uniform and rapid over the time scale of a given test pulse that is used to probe the metric or, in our case, over the time scale over which the photon pairs are generated. In short, we propose an experimental setup similar to that shown in Fig.~\ref{fig:layout} in which a very thin and highly nonlinear medium is pumped by an intense, ultrashort laser pulse. For example, we may use gold or graphene that have very high optical nonlinearity, $n_2$, deposited either in a thin film or in a multilayer structure with an overall thickness that can be of the order of a few nms up to a micron. The pump pulse that excites the nonlinearity should have a rise time that is as short as possible yet larger than the film thickness in order to ensure that there are time transients during which only the leading edge (giving rise to a continuously increasing $n(t)$) or trailing edge (giving rise to a continuously decreasing $n(t)$) overlap with the film.\\
 As an example we choose $\delta\chi(t)=\chi_0\textrm{tanh}(\alpha t)$ with the amplitude $\chi_0$ chosen such that the maximum refractive index variation in the medium is $10^{-2}$ and we study the emission for various rise times $\tau=1/\alpha=1.25$, 2.5 and 5 fs. Results are shown in Fig.~\ref{fig:expansion} for the specific case in which the photons are emitted back to back along the z direction, i.e. perpendicular to the plane of the thin film. We found that this configuration gave an emission that was 2 or more orders more efficient than other configurations e.g. two photons emitted in the forward direction or photons emitted in the x-y plane (data not shown).\\
The first observation is that the model does indeed predict the creation of photon pairs from the time-varying medium thus supporting the idea that our artificial expanding universe interacts with the vacuum by creating new photon states. The emitted photons also show a clear frequency dependence with a peak emission that increases with increasing $\alpha$, i.e. faster changes of rise times $\tau$ of the $n(t)$ lead to higher frequency photons and also higher photon count rates. \\
As can be seen, even if we have chosen a very large $\delta n$, the actual number of photons is extremely small and very unlikely to be detectable. The total photon numbers emitted per second obtained by integrating over the whole frequency range are $N\simeq10^{-2}$, $10^{0}$ and $10^{2}$ for $\tau=5$, 2.5 and 1.25 fs, respectively. These numbers are then further reduced by $\sim15$ orders of magnitude due to the fact that emission is stimulated only for the duration of the $n(t)$ rise time, $\tau$. Reducing the $\delta n$ to more reasonable levels, e.g. $10^{-3}-10^{-4}$ worsens this situation even further. We also verified that changing the specific shape of the $\delta\chi(t)$, e.g. to a Gaussian pulse shape, does not appreciably change this result. However, in the next section we propose a method by which the photon emission mechanism examined here may be resonantly enhanced to experimentally detectable levels.

\section{Resonant enhancement of photon production by periodic contraction-expansion.}
In order to enhance the photon emission whilst maintaining the same dielectric medium geometry shown in Fig.~\ref{fig:layout}, we consider the case in which the laser pump pulse is actually a periodic train of pulses such that we may model the medium response with a 
\begin{equation}
\delta \chi (t,x,y) =  \kappa H(x,y) \cos (\Omega t) 
\end{equation}
where $H(x,y)$ is a function that defines the shape of the film and $\Omega$ is the oscillation period of the laser pulse train and hence also of the medium. Experimentally it is possible to create a train of laser pulses by super-imposing long or even continuous wave laser beams with different frequencies. In the case in which we use frequencies that are harmonics of some fundamental wavelength, it is actually possible to perform Fourier synthesis in the optical domain and create laser pulse trains with arbitrary shapes. The great advantage of this technique, with respect e.g. to resorting to a standard pulsed laser system is that the pulse periodicity may be driven to extremely high values. For example, combining two beams with frequencies $\omega_a$ and $\omega_b$ delivers a waveform with an envelope that oscillates with a periodicity of $\Omega=(\omega_a-\omega_b)$ - if we choose to use a standard Nd:Yag laser (fundamental wavelength, 1064 nm) and its fourth harmonic as the two generating beams, then $\Omega=5.3$ rads/fs which is more than six orders of magnitude larger than what is obtained from standard ``high'' repetition rate lasers that will typically operate at a maximum rate in the range of 100 MHz.\\
  The Fourier transform of $\delta\chi$ determines the properties of the emitted photons. For the specific case in which the film is square with side length $L$, as previously considered we have
\begin{eqnarray}\label{eq_mod}
|\widehat {\delta\chi} (\omega_{1}+\omega_{2}, \mathbf {k_1}+\mathbf {k_2})|^2= \nonumber \\
= \frac{ \pi}{2} \kappa^2
  \delta (\omega_{1}+\omega_{2}-\Omega)  L^6\prod_{j=x,y,z}\left|\textrm{sinc}\left(L\frac{k_{1j}+k_{2j}}{2}\right)\right|^2  
 \end{eqnarray}
\begin{figure}
\centering \includegraphics[width=10cm]{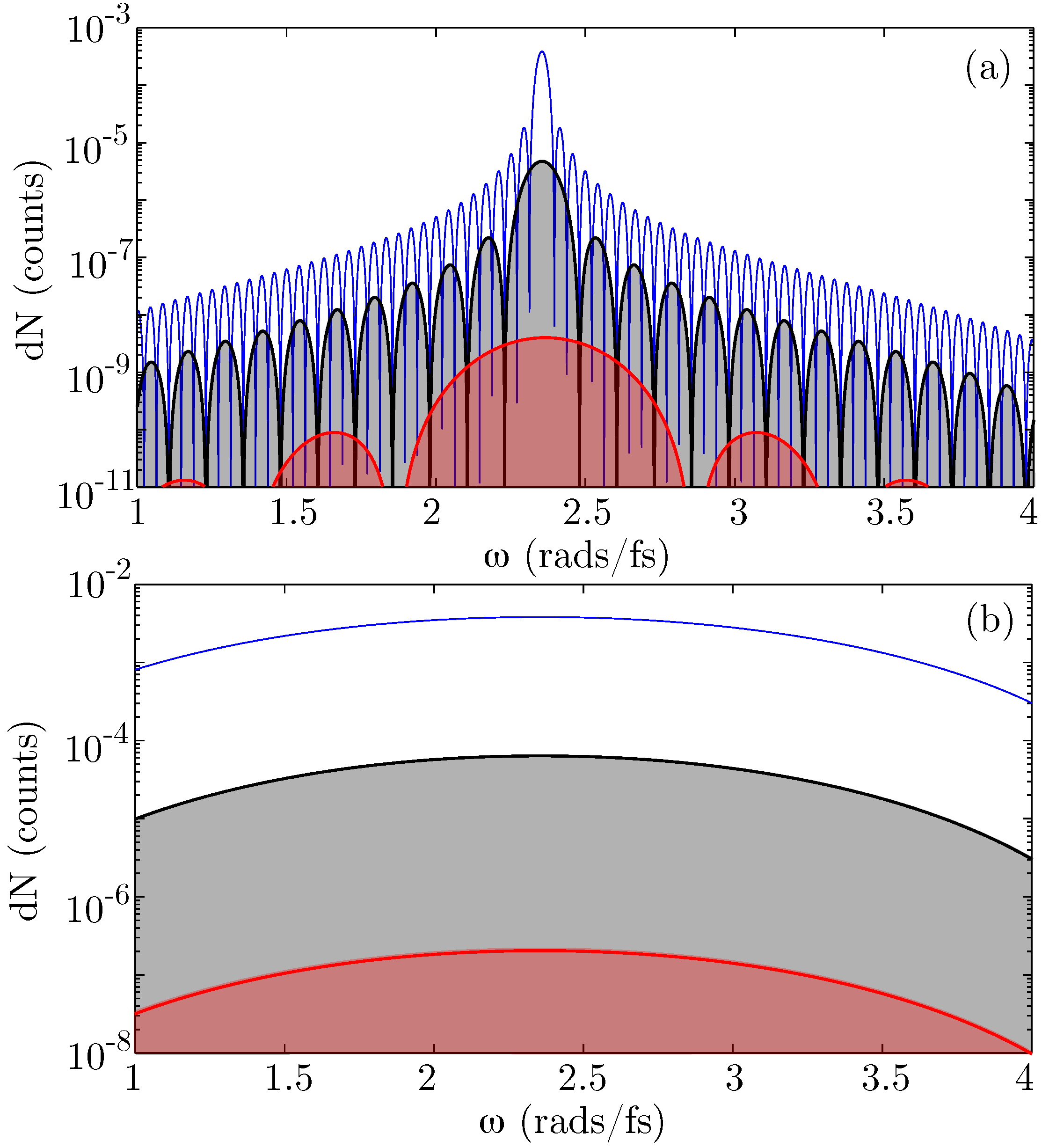}
\caption{\label{fig_b2b}  (a) Number of photon pairs for back to back emission in the x direction (transverse to the thin film). (b) Number of photon pairs for back to back emission in the z direction (perpendicular to the thin film). In each graph, three different square film sizes are considered, with side lengths of 2 (blue line), 6 (black line, grey shaded area) and 20 $\mu$m (red line, red shaded area).  }
\end{figure}
Interestingly, the introduction of a periodic modulation introduces a $\delta$ function that now strongly conditions the emitted frequencies, i.e. resonance with the laser pulse train periodicity limits the photon emission to frequencies given by $\omega_{1}+\omega_{2}=\Omega$. This condition is strongly reminiscent of photon emission by the dynamical Casimir effect, i.e. by periodic modulation of a cavity. In our case there is no cavity, yet the periodicity of the modulation imposes similar resonance conditions.\\
\begin{figure}
\centering \includegraphics[width=12cm]{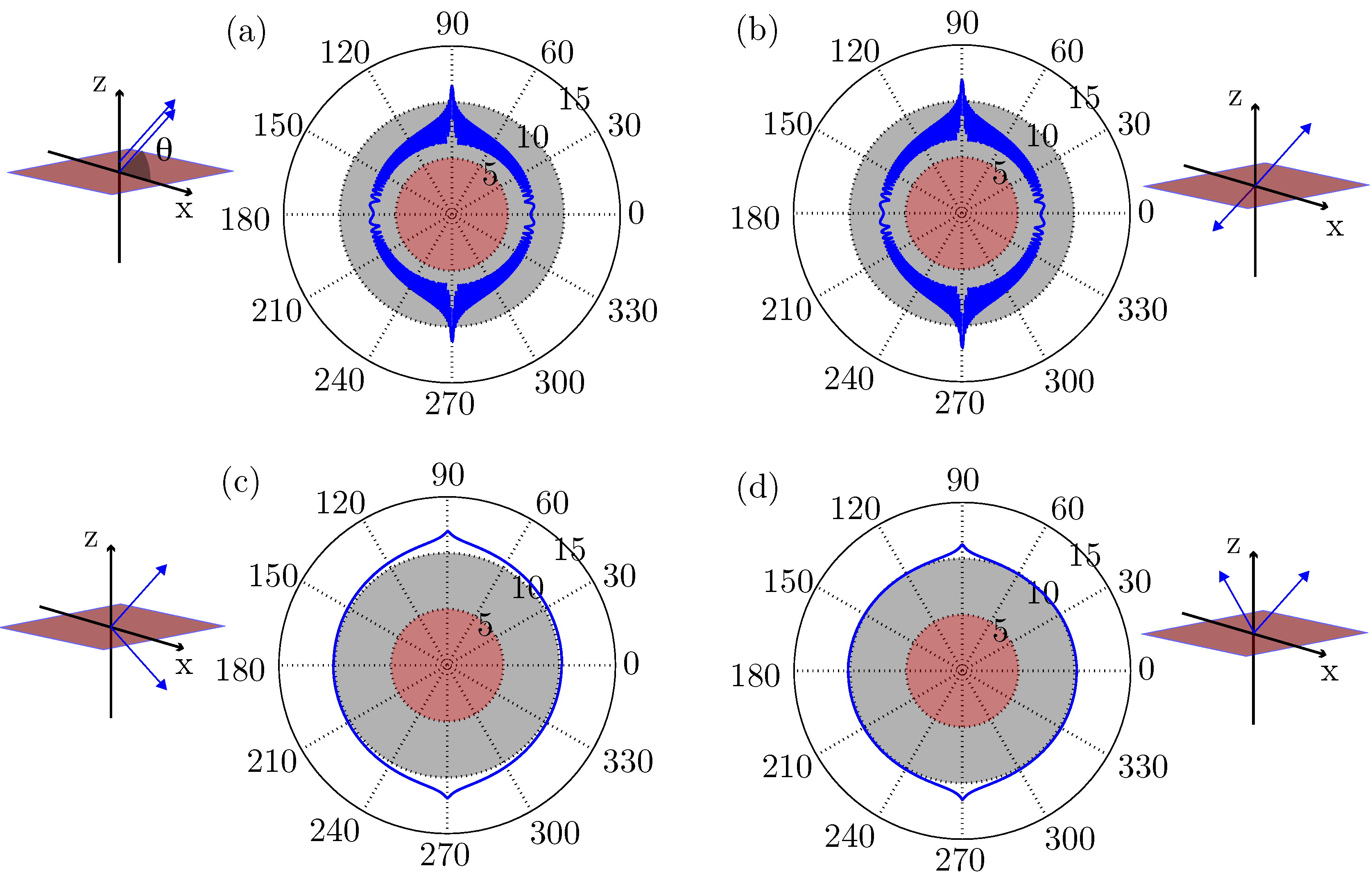}
\caption{\label{fig_polar} Polar plots showing the angular distribution of the photon emission (indicated in total photon numbers/second, log scale). (a) Photon pairs emitted with the same x and z components, (b) Photon pairs with opposite x and z (back to back), (c) Photon pairs with same x and opposite z, (d) Photon pairs with opposite x and same z. }
\end{figure}
Integration of Eq.s~(\ref{eq_hop}) and (\ref{eq_mod}) provide us with the photon number estimates shown in Fig.~\ref{fig_b2b} from a film with thickness 200 nm for the case of back to back emission in the x (or y) direction [Fig.~\ref{fig_b2b}(a)] and z direction  [Fig.~\ref{fig_b2b}(b)]. The different film sizes are considered, i.e. square films with side length $L=2$ (red, shaded curve), 6 (grey, shaded curve) and 20 (blue line) $\mu$m. The modulation frequency is taken to be $\Omega=4.72$ rads/fs.\\ 
As a first observation, the photon numbers are now many orders of magnitude higher with respect to the single $n(t)$-ramp case. The photon counts/second obtained by integrating over the shown spectra are $\sim10^{12}$ and $10^{11}$ for emission in the x and z directions, respectively. If we then consider that we may generate intense, periodic pulse trains as described above with durations of 1-10 nanoseconds, we estimate a total of $10^2-10^3$ photons/pulse. Such photon numbers are well within reach of current photon counting technologies.\\
We note moreover that the emission in the x direction shows clear evidence of a resonance effect induced by the hard-edge boundary of the square film that is acting as a weak cavity of sorts. The larger the film, the narrower the bandwidth of the photon emission peak. Notably, this peak is always centred at the same frequency, $\omega=2.36$ rads/fs that is determined not by the film dimensions but rather, by the dynamical Casimir resonance condition $\omega_1=\omega_2=\Omega/2$, where the condition $\omega_1=\omega_2$ arises from the choice of considering only back to back photon emission. Emission in the z direction is also peaked at the same $\Omega/2$ frequency although the subwavelength dimensions of the film do not enforce any cavity-like effects and hence leads to remarkably broadband emission.\\
In Fig.~\ref{fig_polar} we show the angular emission pattern for four different cases, as indicated by the schematic drawing next to each of the four polar plots. The polar angle is measured such that 0 deg corresponds to emission along the x direction and 90 deg along the z direction.  An interesting observation is the clear angular  pattern that occurs when the photon emission is either such that both photons are emitted exactly in the same direction or exactly back to back: in both cases oscillations appear in the angular emission due to interference between the emission probabilities of the two photons that fall in and out of phase as the angle is changed. Maximum emission for all cases is observed in the z direction.

\section{Analogy with a Gravitational wave spacetime metric.}
In the previous section we analysed the photon production from a medium whose refractive index varies as $n(t) = n_0 + \delta n \cos(\Omega t)$ in the lab frame and the relevant spacetime explicitly takes the form
\begin{equation}
ds^2 = c^2 dt^2 - dx^2[n_0 + \delta n \cos(\omega t)]^2.
\end{equation}
Assuming that $\delta n\ll1$ we find that the null geodesics describing photon trajectories are described  by the equation
\begin{equation}\label{eq_null1}
\frac{dx}{dt} \simeq \frac{c}{n_0} [1-\frac{\delta n}{n_0}\cos(\omega t)].
\end{equation}
At this point it would seem that systems with similar metrics should also have similar quantum emission properties, and now we can use this to look for other interesting structures. \\
The periodically modulated $\delta n$ gives rise to a metric with an oscillating velocity for the 'flow of space', or in other words, a gravitational pull that oscillates periodically. \\
Gravitational waves were first predicted in 1916 by Albert Einstein as ripples in spacetime that propagate in the underlying spacetime, and is thought to be a consequence of the Lorentz invariance of general relativity; the speed of which information about fields travel cannot exceed the speed of light. These ripples are created by numerous binary systems, including those of black holes or neutron stars and carry away energy from the orbit. However, as of this day, there has not been any direct observation of a gravitational wave, despite many years of experimental trial and increasingly sensitive detectors. \\
The interaction of gravitational waves with other particles, e.g. bosons, has attracted significant attention in the past. G. Gibbons first predicted that a gravitational wave will not excite massive particles from the vacuum state \cite{gibbons}, although by imposing adequate boundary conditions e.g. a cavity, particle production may be expected \cite{sorge1,fuentes}. The specific case of interaction with photons has also been considered. Similarly to massive particles, photons were explicitly predicted to not be created by a gravitational wave \cite{montanari} but the presence of certain boundary conditions can lead to the creation of electromagnetic waves, e.g. in the presence of a background magnetic field or plasma \cite{brodin2,brodin1}.\\
Here we consider the case of gravitational wave passing through a limited ``slice'' of space, thus isolating a region similarly to the dielectric film considered above.
So if we choose to look at the case when we have gravitational waves propagating in the z-direction on a flat spacetime background, the resulting metric takes the following form:
\begin{eqnarray}
ds^2 = c^2 dt^2 - dx^2[1+h_+ \cos(\omega(t-\frac{z}{c}))] \nonumber \\
- dy^2[1-h_+ \cos(\omega(t-\frac{z}{c}))] 
- 2 h_\textrm{x} \cos(\omega(t-\frac{z}{c}))dxdy
\end{eqnarray}
where $h_+$ and $h_{\textrm{x}}$ denote the gravitational wave amplitudes and are related to the polarisation of the wave. We now take special interest in the case where only $t$ and $x$ change. The metric then takes the form
\begin{equation}
ds^2 = c^2 dt^2 - dx^2[1+h_+ \cos(\omega t)] 
\end{equation}
\noindent and under the assumption that $h_+\ll1$, we arrive at a similar formula to Eq.~(\ref{eq_null1}) for photon trajectories,
\begin{equation}
\frac{dx}{dt} = \frac{c}{\sqrt{1+h_+ \cos(wt)}} \simeq c(1-\frac{h_+}{2}\cos(\omega t))
\end{equation}
 The similarity to  Eq.~(\ref{eq_null1}) implies that we have a system where light experiences an $x$ coordinate that periodically changes in time in a nearly identical manner. 
 Rather than attempt to derive a full quantum treatment of the problem, here we limit the analysis to demonstrating that gravitational waves may act as amplifiers for electromagnetic waves. This demonstration relies solely on the kinematic analogy of the spacetime contraction created by a gravitational wave and a refractive index change in an optical medium. Nevertheless, this simplified approach allows to show the amplification properties of gravitational waves which must therefore also apply to quantum fluctuations of the electromagnetic field. \\
{ Our treatment follows closely the same approach proposed by Mendon\c{c}a et al., in the context of what has been called ``time refraction'', i.e. the interaction of light with a time varying boundary at a fixed position in space \cite{mendonca1}. By considering a simplified square-shaped wave and imposing continuity conditions on the displacement vector at the time boundaries (i.e. continuity between two immediately successive instants, separated by a sudden contraction or expansion in the metric) it is possible to analytically treat the problem of a gravitational wave induced perturbation accounting for the quantum nature of photons and therefore describing the interaction with the vacuum modes \cite{mendonca2}. The output photon modes are related to the input modes by the standard Bogoliubov transformation
\begin{eqnarray}
\hat{a}_\textrm{out}(k) = \alpha\hat{a}_\textrm{in}(k) - \beta \hat{a}_\textrm{in}^\dagger(-k) \\
\hat{a}_\textrm{out}^\dagger(-k) = \alpha \hat{a}_\textrm{in}^\dagger(-k) - \beta \hat{a}_\textrm{in}(k)
\end{eqnarray}
where $\hat{a}$ and $\hat{a}^\dagger$ are the photon creation and annihiliation operators with Bogoliubov coefficients, $\alpha$ and $\beta$ that for a square shaped contraction followed by a square shaped expansion (i.e. one period of the oscillation)  are given by
\begin{eqnarray}
\alpha = \frac{i}{2\zeta}(\zeta^2+1)  e^{-i \omega_1 t}  \\
\beta =  \frac{i}{2\zeta}(\zeta^2-1) e^{i \omega_1 t}.
\end{eqnarray}
The simple fact that these coefficients are different from zero is sufficient to conclude that indeed photons will be amplified and/or excited out of the vacuum state by a gravitational wave. Clearly a more complete model is required in order to evaluate precisely how efficient the process will be. However, it is possible to gain some insight into the process without resorting to more complicated calculations.\\
A first observation regards the precise geometry required to excite photons out of the vacuum state: the  model adopted here relies on the assumption that the medium, (i) varies uniformly over its entire extent and, (ii) that only one compression or expansion event occurs at a time over the whole sample, i.e. the sample must be shorter in the z direction than half the period of the gravitational wave. This is in keeping with the model and assumptions of the thin film with a periodic refractive index.  If the film or region of spacetime affected by the contraction/expansion is thicker than the oscillation period then the average total effect will go to zero. Another way of seeing this relies on the fact that Maxwell's equations in vacuum are Lorentz invariant: it is always possible to find a reference frame (in vacuum) such that the wave and related spacetime metric is static. In this reference (and hence in all reference frames) there can be no photon production. However, the presence of the medium breaks this invariance and is therefore necessary in order to observe photon emission.   \\
One must therefore limit a region of space along the propagation direction of the gravitational wave, e.g. by adopting a cavity geometry similar to that recently proposed by Sab{\'{i}}n et al. \cite{fuentes}, albeit for phonon creation in a Bose-Einstein condensate. Alternatively, we can follow the same proposal developed above for the optical analogue model and simply use a medium that is transparent and sufficiently extended to support the kHz frequencies one may expect to excite from a gravitational wave but at the same time is confined to subwavelength region of the wave itself. An example in this sense could be to replace the thin nonlinear film used in the optical analogue with a superconducting cable: a back of the envelope calculation can be obtained by simply rescaling the calculation for the optical case. The total photon number scales as $\omega^2\times (L_x\times L_y\times L_z)^2 \times \kappa^2$,  with $\kappa$ the wave amplitude. We predict a photon pair created every $\sim10$ minutes if for example we consider a gravitational wave with $\omega\sim100$ kHz, $\kappa\sim 10^{-20}$, and  $L_x=10^4$ m (length of cable), $L_y=L_z=0.1$ m (diameter of cable).\\
We underline that this is just a rough estimate based on our analogue model but we still expect the scaling dependencies to be correct and the overall estimate not too far from truth.  This would of course be a large scale experiment and a detailed analysis and discussion of this is beyond the scope of the present work. The point of this example is simply to underline possible future avenues and elaborate on how the study of analogue models can give some insight into cosmological phenomena.

\section{Analogy with parametric amplification and the dynamical Casimir effect}

The term ``dynamical Casimir effect'' was first coined by Yablanovitch  in reference to the production of photons from an accelerated or sudden, non-adiabatically changing medium \cite{yablanovitch}. The same terminology now typically denotes the creation of entangled quanta, excited out of the vacuum state by means of a periodically (in time) oscillating boundary condition \cite{moore,fulling,lambrechtDCE}.  It was on the basis of this model that an effect analogous to the DCE was observed experimentally for the first time in oscillating superconducting circuits \cite{delsing,DCE-finland}. It has also been noted that in this context an oscillating cavity effectively behaves as a parametric amplifier and the analogy between the two systems has been analysed in detail \cite{lambrecht,johansson,nori}.\\
In the system analysed here we do not explicitly have a cavity of any form. In our periodically oscillating system in which it is not the boundary conditions, e.g. the extremities of the medium that are oscillating but rather it is the medium as a whole that oscillates in time. This is very similar to other realisations of the DCE that have been proposed based on static mirrors that confine a time-varying medium \cite{lambrecht,FaccioEPL,DCE-finland,DCE_BEC}. In the optical analogue, the resonant enhancement from the periodic modulation alone is sufficient to render the photons detectable.
As for the DCE, the periodic modulation described here also bears a close resemblance to parametric amplification in nonlinear optical media. By pumping a nonlinear crystal with an intense beam two entangled  photons, called signal and idler are excited out of the vacuum state \cite{boyd} through a mechanism that relies on the periodic nature of the pump beam. 
So there is a strong similarity and connection between the effects studied here and standard parametric amplification. Indeed, the energy conservation rules are the same ($\omega_1+\omega_2=\Omega$) where $\Omega$ can be either the parametric oscillator frequency or, in our case, the medium oscillation frequency. In this sense, the system we are analysing is most certainly a form of parametric amplification. This is true in general for all systems that mimic a dynamical Casimir effect. However, the difference with respect to a traditional parametric amplifier employed in nonlinear optics experiments lies in the specific details of the boundary conditions. A typical nonlinear optics amplifier will be made of a long crystal, i.e. a crystal that is many wavelengths long. The distance-averaged oscillation inside the crystal is therefore zero: the medium as whole is not oscillating in time. On the other hand, in our setting the medium has a sub-wavelength thickness and averaging over the whole propagation distance, still leaves us with medium that is oscillating as a whole, in time. This is what allows us to write out a spacetime metric for the system that includes a term with $n=n(t)$ and thus, on the basis of this, draw an analogue with similar cosmological phenomena [this is not possible for a long standard NLO crystal as this would require $n=n(z-vt)$].
In closing, we also note that {\emph {in vacuum}} the Lorentz-invariance of Maxwell's equations ensures that we may always find a reference frame within which the perturbation or overall system does not change in time. Photon production will not occur in this reference frame and we must therefore conclude that it will not occur in any other frame either. However, the presence of the medium breaks the Lorentz invariance: there is no reference frame in which both the medium and the refractive index oscillation are stationary. This is the key to the effect in the optical analogue and also to the production of kHz waves in the superconducting wire example we gave above.

\section{Concluding remarks.}
In conclusion, based on considerations dealing with the optical analogue, we have shown that photon creation may occur through a periodic modulation of a medium that is localised in the transverse directions and strongly subwavelength in the longitudinal coordinate.} Optical nonlinearities in very thin transparent optical media excited by a laser pulse or a train of laser pulses will give rise to a refractive index that varies in time, uniformly across the sample. Such a geometry can model a cosmological expansion or contraction and can thus be used to study details of how photons may be excited in similar spacetime metrics. A single contraction/expansion appears to deliver a very weak flux of photon pairs. However, the flux may be considerably enhanced by performing a periodic modulation, in which case a resonance is observed in the emitted photon frequencies corresponding to half the modulation frequency. The optical spacetime metric bears a similarity to a gravitational wave which will also amplify electromagnetic waves. However, this will occur only if  boundary conditions are imposed, e.g. such that the interaction of the gravitational wave with the electromagnetic vacuum is limited to a region that is smaller than the gravitational perturbation wavelength. It seems that direct detection of actual gravitational waves based upon photon excitation in the kHz range using such an a approach could be feasible, notwithstanding the extreme weakness of the gravitational waves themselves, which in turn implies that the actual photon counts will be very low. The analogue dielectric medium model proposed here however, does allow a greater degree of control over the experimental conditions and thus provides an intriguing possibility to study the quantum features of these and other cosmological effects using laboratory based experiments.

\section*{Acknowledgements}
D.F. acknowledges discussions with I. Fuentes and financial support from the European Research Council under the European Union’s Seventh Framework Programme (FP/2007-2013)/ERC GA 306559 and EPSRC (UK, Grant EP/J00443X/1).

\section*{References}

\bibliography{bibliography}{}
\bibliographystyle{unsrt}

\end{document}